\documentclass[aps,amssymb,twocolumn]{revtex4}

\usepackage{graphicx}
\usepackage{amsmath}
\usepackage{epsfig}
\usepackage{bm}

\newcommand{\be}{\begin{equation}}
\newcommand{\ee}{\end{equation}}
\newcommand{\beq}{\begin{eqnarray}}
\newcommand{\eeq}{\end{eqnarray}}

\makeatletter \leftline{\epsfbox{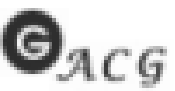}}
 \leftline
{\footnotesize{\textbf{\textsf{GACG-04/11}}}}

\begin{document}

\title{The geodesic structure of the Schwarzschild Anti-de Sitter black hole}
\author{Norman Cruz }
 \email{ncruz@lauca.usach.cl}
\author{ Marco Olivares }
 \email{marcofisica@gacg.cl}
\author{ J.R.Villanueva }
 \email{jrvillanueva@gacg.cl}

\affiliation{\it Departamento de F\'{\i}sica, Facultad de Ciencia,
Universidad de Santiago de Chile, Casilla 307, Santiago 2, Chile}

\date{\today}

\begin{abstract}
In the present work we found the geodesic structure of an AdS
black hole. By means of a detailed analyze of the corresponding
effective potentials for particles and photon, we found  all the
possible motions which are allowed by the energy levels. Radial
and non radial trajectories were exactly evaluated for both
geodesics. The founded orbits were plotted in order to have a
direct visualization of the allowed motions.  We show that the
geodesic structure of this black hole presents new type of motions
not allowed by the Schwarzschild spacetime.
\end{abstract}

\maketitle

\section{INTRODUCTION}

The presence of a vacuum energy (cosmological constant) has been
considered in theoretical models related to unification as
superstring and to cosmology and astrophysics. A possitive
cosmological constant is supported by the measurements of distant
SNe Ia (at $ z\sim 1$), which indicate that the expansion of the
Universe is in accelaration \cite{Peetal}. Other tests of the
standard model, including spacetime geometry, galaxy peculiar
velocities, structure formation, and early Universe physics,
supports, in many of these cases, a flat Universe model with the
presence of a cosmological constant \cite{Pe2}.

Of course, the above evidences have motivated to consider
spherical symmetric spacetimes with non zero vacuum energy in
order to study the well known effects predicted by General
Relativity for planetary orbits and massless  particles. This
study implies to determine the geodesic structure of Kottler
spacetimes \cite{kotler}. Timelike geodesics  for positive
cosmological constant were investigated in \cite{Jaklitsch},
using only the method of an effective potential in order to found
the conditions for the existence of bound orbits. The analyze of
the effective  potential  for radial null geodesic in
Reissner-Nordstrom-de Sitter and Kerr-de Sitter spacetime was
realized in \cite{Stuchlik}. Podolsky \cite{podolsky} investigated
all possible geodesic motions for extreme  Schwarzschild-de Sitter
spacetime.

Classical test of general relativity such as the bending of light
was examined by Lake \cite{Lake}, founding that the cosmological
constant produces no change in this effect. Exact solutions in
closed analytic form for geodesic motion in Kottler spacetime were
found very recently by Kraniotis {\it et al}  \cite{K-W}. The
exact solutions of the timelike geodesic were used to evaluate the
perihelion precession of the planet Mercury. The motion of massive
particles in the Kerr and Kerr-(anti) de Sitter  gravitational
fields was investigated in \cite{Kraniotis}, where the geodesic
equations are derived, by solving the Hamilton-Jacobi partial
differential equation.

The main purpose of this article is to find exact solutions for
time-like and null  geodesic of a Schwarzschild Anti-de Sitter
spacetime. We have restricted to this spacetime since our
objectives are: i) to investigate by means of a detailed analyze
of the effective potentials,  all the possible movements which are
allow by the energy levels, ii) to find the exact solutions
describing the trajectories of massive and null particles, and
iii) to plot the founded orbits in order to have  a direct
visualization of the allowed motions.

In section $2$, we derive the geodesic equations of motion using
the variational problem  associated to this metric. We find the
equivalent one dimensional problem with the respective effective
potential.

In section $3$ the effective potential is analyzed in order to
determine the possible motions  of massive particles. The exact
solutions are calculate for the allowed motions. In particular, we
show the exact solution in the proper time for a radial geodesic.
Planetary orbits are integrated to obtain the polar equation in
term of incomplete integrals of Jacobi. Using an elementary
derivation the advance of perihelion is evaluated. We also gives
an exact solution for planetary orbits which could end into the
event horizon in terms of elementary functions.

In section $4$ null geodesics are analyzed in terms of the one
dimensional effective potential. Explicit exact solutions are
found for radial geodesics in terms of the proper and coordinate
times. Bounded and unbounded geodesics are integrated in terms of
Jacobian elliptic integrals. The bounded geodesics do not exist in
the Schwarzschild spacetime.

In section $5$ we summarize and discuss the results
founded.

\section{GEODESICS}

\noindent The metric for a static spherically symmetric spacetime
with mass $M$ and a negative cosmological constant $\Lambda =
-3/\ell^{2} $ is

\be
ds^{2}=-f(r)dt^{2}+\frac{dr^{2}}{f(r)}+r^{2}(d\theta^{2}+sin^{2}\theta
d\phi^{2}), \label{1.1} \ee

\noindent where the lapse function, $f(r)$, is given by

\be f(r)=1-\frac{2M}{r}+\frac{r^{2}}{\ell^{2}}, \label{1.2}\ee

\noindent and the coordinates are defined such that $-\infty \leq
t \leq +\infty$, $r \geq 0$, $0 \leq \theta \leq \pi$ and $0 \leq
\phi \leq 2\pi$.

\noindent The lapse function vanished at the zeros of the cubic
equation

\be r^{3}+\ell^{2}r-2M\ell^{2}=0. \label{1.3}\ee

\noindent The only real root of this equation is (see appendix A
for more details)

\be
r_{+}=\frac{2}{3}\sqrt{3}\ell\sinh \left[ \frac{1}{3}\sinh ^{-1}\left( 3%
\sqrt{3}\frac{M}{\ell}\right) \right] \label{1.4}\ee

Expanding $r_{+}$ in term of $M$ with $1/\ell^{2} \ll M^{2}/9$, we
obtain

\be r_{+}\approx 2M-\frac{8M^{3}}{\ell^{2}}+...,\label{1.41}\ee

The event horizon of the {\it AdS} black hole is lower than the
Schwarzschild event horizon, $r_{+Sch}=2M$.

\noindent In order to find the geodesics structure of the
spacetime  described by (\ref{1.1}), we solve the Euler-Lagrange
equations for the variational problem associated to this metric
(see \cite{Adler}, for instance). The corresponding Lagrangian is

\be
\mathcal{L}=-f(r)\dot{t}^{2}+f^{-1}(r)\dot{r}^{2}+r^{2}(\dot{\theta}%
^{2}+sin^{2}\theta \dot\phi), \label{1.6} \ee

\noindent where the dots represents the derivative with respect to
the affine parameter $\tau$, along the geodesic.

\noindent The equations of motion are

\be \dot{\Pi}_{q} - \frac{\partial \mathcal{L}}{\partial q} = 0,
\label{1.7} \ee
\bigskip

\noindent where $\Pi_{q} = \partial \mathcal{L}/\partial \dot{q}$
is the momentum conjugate to coordinate $q$.

\noindent Since the Lagrangian is independent of ($t, \phi$) the
corresponding conjugate momenta are conserved, therefore

\be \Pi_{t} = -(1-\frac{2M}{r}+\frac{r^{2}}{\ell^{2}}) \dot{t} = -
E, \label{1.8}\ee

\noindent and

\be \Pi_{\phi} = r^{2}sin^{2}\theta \dot{\phi} = L, \label{1.9}\ee

\noindent where $E$ and $L$ are constant of motion.

\noindent Now, from the equation of motion for $\theta$, we have

\be \frac{d(r^{2}\dot\theta)}{d\tau}=r^{2}sin\theta cos\theta
\dot\phi^{2},. \label{1.10} \ee

\noindent For simplicity, let us choose the initial condition
$\theta = \pi/2$ and $\dot\theta = 0$. Then, from the last
equation we find that $\ddot\theta = 0$. This means that the
motion is confined to the plane $\theta = \pi/2$, which is
characteristic of the \textit{central fields}. With this election
equation (\ref{1.9}) becomes

\be \Pi_{\phi} = r^{2}\dot{\phi} = L, \label{1.11}\ee

\noindent and, from (\ref{1.8})and (\ref{1.11}), the Lagrangian
(\ref{1.6}) can be write in the following form

\be 2\mathcal{L}\equiv
h=\frac{E^{2}}{1-\frac{2M}{r}+\frac{r^{2}}{\ell^{2}}}-\frac{\dot r^{2} }{1-%
\frac{2M}{r}+\frac{r^{2}}{\ell^{2}}}-\frac{L^{2}}{r^{2}}.
\label{1.12} \ee

\noindent By normalization, we shall consider that $h = 1$ for
massive particles and $h = 0$ for photons . We solve the above
equation for $\dot r^{2}$ in order to obtain the \textit{radial
equation}, which allow us to characterize possible moments of test
particles without and explicit solution of the equation of motion
in the invariant plane

\be
\dot r^{2}= E^{2}-\left(1-\frac{2M}{r}+\frac{r^{2}}{\ell^{2}}\right)\left(h+%
\frac{L^{2}}{r^{2}}\right). \label{1.13}\ee

\noindent It is useful to rewrite (\ref{1.13}) as the equation of
motion of a one dimensional problem

\be \dot r^{2}= E^{2}-V^{2}_{eff}. \label{1.14} \ee

\noindent where $V^{2}_{eff}$ define an \textit{effective
potential}

\be
V^{2}_{eff}=\left(1-\frac{2M}{r}+\frac{r^{2}}{\ell^{2}}\right)\left(h+\frac{%
L^{2}}{r^{2}}\right), \label{1.15} \ee

\section{TIME-LIKE GEODESICS}

\noindent For time-like geodesic $h=1$ and the \textit{effective
potential} becomes

\begin{equation}
V^{2}_{eff}=\left(1-\frac{2M}{r}+\frac{r^{2}}{\ell^{2}}\right)\left(1+\frac{%
L^{2}}{r^{2}}\right).\label{2.1}
\end{equation}
\medskip

\noindent Using this effective potential, we  solve the
\textit{radial equation} (\ref{1.13}) for radial and non-radial
particles.

\begin{figure}[!h]
  \begin{center}
    \includegraphics[width=85mm]{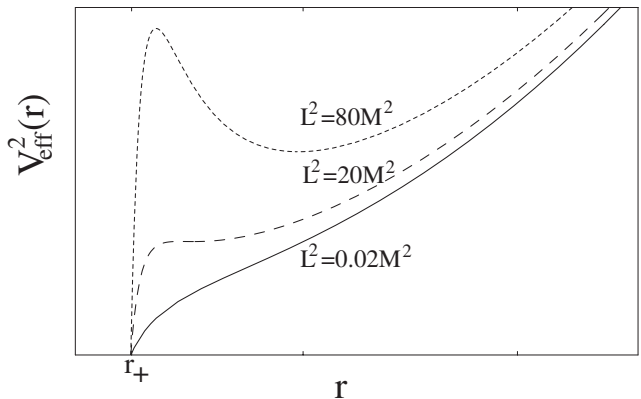}
  \end{center}
  \caption{The figure shows the evolution of the effective potential for non-radial particles at different
  values of the constant of motion $L$.}
  \label{fig:fig.2}
\end{figure}

\subsection{Radial geodesics}

\begin{figure}[!h]
  \begin{center}
    \includegraphics[width=85mm]{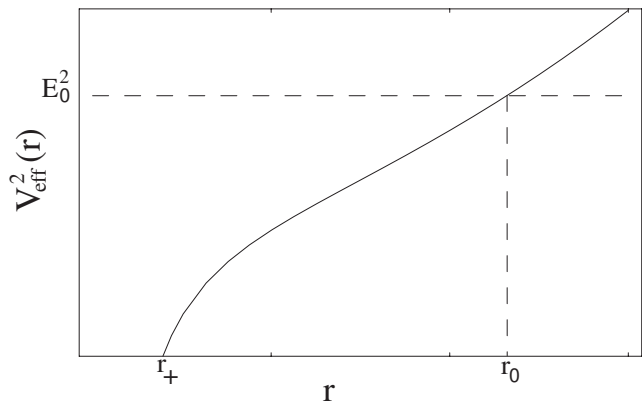}
  \end{center}
  \caption{The figure shows the  effective potential for radial particles. }
  \label{fig:fig.1}
\end{figure}

\noindent The radial geodesic correspond to the motion of
particles without angular momentum $L = 0$, which are falling from
rest, in a finite distance, to the center. In this case, the
effective potential becomes

\begin{equation}
V^{2}_{eff}=\left(1-\frac{2M}{r}+\frac{r^{2}}{\ell^{2}}\right).
\label{2.1.1}
\end{equation}

\noindent which is showed in FIG.\ref{fig:fig.1}. Therefore, the
radial equation is

\be \dot r^{2}=
E^{2}-\left(1-\frac{2M}{r}+\frac{r^{2}}{\ell^{2}}\right).
\label{2.1.1.0}\ee

\noindent In the elliptic curve representing $V^{2}_{eff}$
particles always plunges into the horizon from an upper distance
determined by the constant of motion $E$. This fact is due to
attractive force generated by $\Lambda < 0$. In order to find
exact solutions we solve, by simplicity, the case when two of the
roots of the cubic equation are equals \footnote{This mean that
the solution of the elliptic curve cannot be a elliptic
function.}.
The corresponding energy is $E^{2}_{0} = 1 + 3\sqrt[3]{%
\frac{M^{2}}{\ell^{2}}}$. For this value of $E$, we obtain that
$r_{0}=2\sqrt[3]{M\ell^{2}}$. The constant of motion $E$ in terms
of $r_{0}$ yields

\be E^{2}_{0} = 1 +
\frac{3}{4\ell^{2}}r_{0}^{2}.\label{2.1.1.1}\ee

Replacing this values in  the radial equation (\ref{2.1.1.0}), we
obtain

\begin{equation}
-\frac{dr}{d\tau}=\frac{1}{\ell}\left(r+\frac{r_{0}}{2}\right)\sqrt{\frac{%
r_{0}-r}{r}}.  \label{2.1.2}
\end{equation}
\bigskip

\noindent Making the change of variable $r=r_{0} cos^{2}\eta/2$,
the above equation yields

\begin{equation}
\frac{d\eta}{d\tau}=\frac{1}{\ell}\frac{2+\cos\eta}{1+\cos\eta}.
\label{2.1.3}
\end{equation}

\noindent A forward integration of this last equation yields

\be \tau(\eta)=2\ell\left[\frac{\eta}{2}+
\frac{\sqrt{3}}{3}\arctan\frac{\sqrt{3}}{3}\tan\left(\frac{\eta}{2}\right)\right]
\label{2.1.4}\ee

In term of the variable $r$, ( $\eta_{0}=0$ if $r=r_{0}$ and, from
(\ref{2.1.4}), $\tau_{0}=0$)

\be \tau(r)=2\ell\left[\arccos\left(\frac{r}{r_{0}}\right)+
\frac{\sqrt{3}}{3}\arctan\frac{\sqrt{3}}{3}\frac{\sqrt{r_{0}-r}}{r}\right]
\label{2.1.5}\ee

\begin{figure}[!h]
  \begin{center}
    \includegraphics[width=75mm]{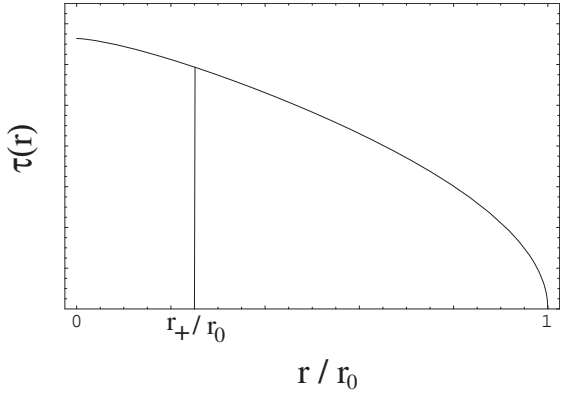}
  \end{center}
  \caption{The figure shows the proper time $\tau$ as a function of $r/r_{0}$.}
  \label{fig:fig.12}
\end{figure}

\noindent In FIG.\ref{fig:fig.12} equation (\ref{2.1.5}) is
plotted. The particle falls to the horizon in a finite proper time
lower than in the Schwarzschild case.

Now, we shall see that the situation is very different when
consider the equation of the trajectory in coordinate time $t$.
The equation to be integrated in order to obtain $t$ is (cf.
equation (\ref{1.8}) and (\ref{2.1.2}))

\be
\frac{dt}{dr}=-\ell^{3}E\frac{r}{(r^{2}+r_{+}r+\rho^{2})(r-r_{+})
(r+\frac{r_{0}}{2})\sqrt{\frac{r_{0}-r}{r}}},\label{2.1.6}\ee

\noindent Using the substitution $r=r_{0}\cos^{2}\eta/2$, this
equation gives

\be t(\eta)=A[B\tau(\eta)+C H(\eta)+D
\eta-I(\eta)],\label{2.1.61}\ee

\noindent where

\beq &&A=\frac{\ell^{3} E}{r_{0}+2r_{+}},\qquad
B=\frac{r_{0}}{\rho^{2}+\frac{r^{2}_{0}}{4}-\frac{r_{0}r_{+}}{2}},\nonumber\\
&& C=\frac{2r_{+}}{\rho^{2}+2r_{+}},\qquad
D=B+C,\label{2.1.62}\eeq

\noindent and

\be
H(\eta)=\eta+\cot\frac{\eta_{+}}{2}\ln\left|\frac{\tan\frac{\eta_{+}}{2}+\tan\frac{\eta}{2}}
{\tan\frac{\eta_{+}}{2}-\tan\frac{\eta}{2}}\right|,\label{2.1.63}\ee

\be
I(\eta)=\int^{\eta}_{0}\frac{a\cos^{2}\frac{\eta}{2}+b}{r^{2}_{0}\cos^{4}\frac{\eta}{2}+r_{0}r_{+}\cos^{2}\frac{\eta}{2}
+\rho^{2}}d\eta.\label{2.1.64}\ee

\noindent Also, $\tau(\eta)$, which is given by (\ref{2.1.4}),
together with the integral $I(\eta)$ are well defined when $\eta
\rightarrow \eta_{+}$. Moreover, in these limit, the function
$H(\eta)$ goes to infinity, which implies that the coordinate
time, $t$, goes to infinity too. This physical result is in accord
with the Schwarzschild case.

\subsection{The bound orbits}

\noindent In this case we have that $L \neq 0$, and the
\textit{radial equation} takes the form

\begin{equation}
\dot r^{2}= E^{2}-\left(1-\frac{2M}{r}+\frac{r^{2}}{\ell^{2}}\right)\left(1+%
\frac{L^{2}}{r^{2}}\right).  \label{2.1.7}
\end{equation}
\bigskip

\noindent In FIG.\ref{fig:fig.2} the effective potential has been
plotted for non-radial particles when $L^{2}>11.25M^{2}$. The
following orbits are allowed depending on the value of the
constant $E$

1.- If $E^{2}=E^{2}_{I}$, the particle can orbit in a stable
circular orbit at $r=r_{c}$. The other possible orbit correspond
to a particle at the point $B$ which plunges into the singularity.
\smallskip

2.- If $E^{2}_{I}<E^{2}<E^{2}_{II}$, the particle orbit on a bound
orbit in the range $r_{P}<r<r_{A}$, where $r_{P}$ and $r_{A}$ are
the perihelion and aphelion distance, respectively. This orbits
will be calling {\it orbits of the first kind}. The other possible
orbit correspond to a particle at the point $D$ which plunges into
the singularity.
\smallskip

3.- If $E^{2}=E^{2}_{II}$, the particle can orbit in a unstable
circular orbit at $r=F$. What is more likely is that a particle is
such type of orbit will recede from $r=F$ to either $r=G$ or the
singularity at $r=0$.
\smallskip

\begin{figure}[!h]
  \begin{center}
    \includegraphics[width=85mm]{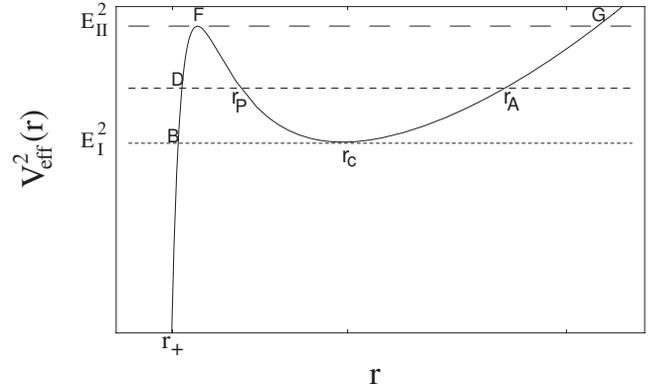}
  \end{center}
  \caption{The figure shows the effective potential for non-radial particles when $L^{2}>11.25M^{2}$.
  The horizontal lines represent different labels of the constant of motion $E^{2}$.}
  \label{fig:fig.2}
\end{figure}

\noindent Is possible to obtain the orbits equations  using
equation (\ref{1.11}) in  (\ref{2.1.7}) and making the change of
variable $u = r^{-1}$, we obtain

\be \left(u\frac{du}{d\phi}\right)^{2}=2M{\it P}(u), \label{2.1.8}
\ee

\noindent where the polynomial ${\it P}(u)$ is given by

\be {\it P}(u)=u^{5}-
\frac{1}{2M}u^{4}+\frac{1}{L^{2}}u^{3}+\frac{\varepsilon^{2}}{2ML^{2}}u^{2}-\frac{1}{2ML^{2}\ell^{2}},
\label{2.1.9}\ee

\noindent and the parameter $\varepsilon^{2}$ is defined by the
relation

\be \varepsilon^{2} \equiv E^{2}-1-\frac{L^{2}}{\ell^{2}} \equiv
E^{2}-E_{\Lambda}^{2}.\label{2.1.10}\ee

\noindent As we see, from FIG.\ref{fig:fig.2}, bound orbits can
exist in an Schwarzschild AdS space-time. This fact is imposed by
the condition $\varepsilon^{2} >0$.

\noindent Since ${\it P}(u)$ is a $5$ degrees polynomial, our
analyze take an easy form if we consider that this polynomial
posses two identical negative roots, $u_{4}=u_{5}=-\sigma$.
Therefore, we have

\be {\it P}(u)=(u+\sigma)^{2}g(u),\label{2.1.11}\ee

\noindent where $g(u)$ is a cubic polynomial, which posses the
information of the physical range of interest. This polynomial is
the key for the study of the different kind of orbits that we
shall pass to analyze. In term of $g(u)$, (\ref{2.1.8}) is
integrated, yielding

\begin{widetext}
\be \pm \Delta
\phi=\frac{1}{\sqrt{2M}}\int^{u}_{u_{0}}\frac{du'}{\sqrt{g(u')}}
-\frac{\sigma}{\sqrt{2M}}\int^{u}_{u_{0}}\frac{du'}{(u'+\sigma)\sqrt{g(u')}}.
\label{2.1.12}\ee
\end{widetext}

\subsubsection{\textbf{Orbits Of The First Kind}}

The orbits of the first kind represents bound geodesics, with two
extreme values. These type of orbits are allowed when $E$ and $L$
satisfy $E<1$ and $L^{2}>11.25M^{2}$. We can identify $3$ class of
orbits of the first kind: Planetary orbits, asymptotic orbits and
circular orbits.
\medskip

(i)\underline{\textbf{Planetary orbits}}
\medskip

\noindent The planetary orbits are constrain to oscillate between
an aphelion and a perihelion. These distances correspond to the
roots of $g(u)$, therefore, the another root is also real and
positive, which shall must correspond to the aphelion distance at
the continuation of the planetary orbit. In order to obtain these
root explicitly, we make the substitution

\be u=\frac{1+e\cos\chi}{R},\label{2.1.11.0}\ee

\noindent where $e$ represent the eccentricity of the orbits, and
$R$ is the {\it latus rectum}. Therefore, solving the equations
(\ref{B8}-\ref{B10}), we find

\be u_{1}=\frac{1-e}{R},\qquad u_{2}=\frac{1+e}{R},\qquad
u_{3}=\frac{1}{2M}-\frac{2}{R}+2\sigma. \label{2.1.11.1}\ee

\noindent where $u_{1}$ correspond to the inverse of the aphelion
distance ($\chi=\pi$), $u_{2}$ is the inverse of the perihelion
distance ($\chi=0$) and $u_{3}$ is the inverse of an aphelion
distance for the orbits of second kind, which we shall study in
the next section. The  physical range in this case is $r>r_{3}$
(i.e, $u_{3}>u$). Therefore, the cubic polynomial $g(u)$ can be
rewrite as $g(u)=(u-u_{1})(u-u_{2})(u_{3}-u)$. Thus, the equation
(\ref{2.1.12}) becomes

\begin{widetext}
\be \pm \Delta \phi=\frac{1}{A}\int^{\chi}_{\pi}\frac{-d\chi}
{\sqrt{1-{\it k}^{2}\sin^{2}(\frac{\pi}{2}-
\frac{\chi}{2})}}-\frac{\nu n^{2}}{2 e A}
\int^{\chi}_{\pi}\frac{-d\chi}{(1+n^{2}\sin^{2}(\frac{\pi}{2}-\frac{\chi}{2}))\sqrt{1-{\it%
k}^{2}\sin^{2}(\frac{\pi}{2}-\frac{\chi}{2})}}.
\label{2.1.11.2}\ee \end{widetext}

\noindent where

\be A^{2}=1-6\mu +2\mu e+4\mu\nu, \label{ofkpl1}\ee

\be{\it k}^{2}=\frac{4\mu e}{A^{2}},\label{ofkpl2}\ee

\be n^{2}=\frac{2e}{1-e+\nu}, \label{ofkpl3}\ee

\noindent and

\be \mu=\frac{M}{R} \qquad \nu=\sigma R.\label{ofkpl4}\ee

\noindent In term of incomplete integrals of Jacobi, the above
solution can be rewrite as

\begin{equation}
\phi = \frac{2}{A}\left[F\left({\it k},\frac{\pi}{2} -
\frac{\chi}{2}\right)-\frac{\nu n^{2}}{2 e} \Pi\left(n,{\it
k},\frac{\pi}{2} - \frac{\chi}{2}\right)\right]. \label{2.1.11.3}
\end{equation}

\medskip

(i.1)\underline{Advance of the perihelion}.
\medskip

We use the elementary derivation of the advance of perihelion of a
planetary orbit given in \cite{Cornbleet} for the Schwarzschild
solution. In our case, the advance of the perihelion is obtain
comparing a keplerian ellipse in a Lorentzian coordinates with one
in a Schwarzschild Anti-de Sitter coordinates. The relevant
relation communicating the two ellipse is the areal constant of
Kepler's second law. In unperturbed Lorentz coordinates the line
element is given by

\be ds^{2}=-dt^{2}+dr^{2}+r^{2}(d\theta^{2}+\sin^{2}\theta
d\phi^{2}).\label{ap1} \ee

The Schwarzschild Anti-de Sitter (SAdS) gravitational field, given
by equation (\ref{1.2}), allow to find the following
transformation of the coordinates, $r$ and $t$, in the binomial
approximation

\be dt'=(1-\frac{M}{r}+\frac{r^{2}}{2\ell^{2}})dt,\label{ap2} \ee

\be dr'=(1+\frac{M}{r}-\frac{r^{2}}{2\ell^{2}})dr.\label{ap3} \ee

We consider two elliptical orbits, one the classical Kepler orbit
in $r$, $t$ space and a {\it Schwarzschild Anti-de Sitter} orbit
in an $r'$, $t'$ space.  In the Lorentz space we have

\be dA=\int_{0}^{\rho}r dr d\phi,\label{ap4} \ee

and hence the Kepler second law

\be \frac{dA}{dt}=\frac{1}{2}\rho^{2}\frac{d\phi}{dt}.\label{ap5}
\ee

In the Schwarzschild Anti-de Sitter situation however, we have

\be dA'=\int_{0}^{\rho}r dr' d\phi,\label{ap6} \ee

\noindent with $dr'$ given by equation (\ref{ap3}). Therefore,
applied wherever necessary the binomial approximation and the
transformation of coordinate, we obtain

\be \frac{d
A'}{dt'}=\frac{\rho^{2}}{2}\left(1+\frac{3M}{\rho}+\frac{2M^{2}}{\rho^{2}}-\frac{3\rho^{2}}{4\ell^{2}}
-\frac{5M\rho}{4\ell^{2}}+...\right)\frac{d\phi}{dt}.
\label{ap7}\ee

 Applying all of this increasing for a single orbit

\be \Delta
\phi'=\int_{0}^{2\pi}\left(1+\frac{3M}{\rho}+\frac{2M^{2}}{\rho^{2}}
-\frac{3\rho^{2}}{4\ell^{2}}-\frac{5M\rho}{4\ell^{2}}+...\right)d\phi.
\label{ap8}\ee

For an ellipse $\rho=\frac{R}{1+e\cos\phi}$ where $e$ is the
eccentricity and $R$ is the {\it latus rectum}. Therefore,
applying the binomial approximation, we obtain

\begin{equation}
\Delta \phi' \approx 2\pi+\frac{6\pi M}{R}+\frac{4\pi
M^{2}}{R^{2}}-\frac{3\pi R^{2}}{2\ell^{2}}-\frac{5\pi
MR}{2\ell^{2}}+...,
\end{equation}

the classical advance of perihelion is recuperated for zero
cosmological constant ($\ell \rightarrow \infty$). The two last
terms are the corrections due to a negative cosmological constant.

(ii)\underline{\textbf{Asymptotic orbits}}
\medskip

\noindent This case occur when two real roots of $g(u)$ are
coincident and positive, namely $u_{2}=u_{3}$, and also this roots
are placed in the extremal of the effective potential, such that
the energy is characterized by $E^{2}_{II}$. The substitution is

\be u=\frac{1+e\cos\chi}{R}.\label{ofkasy1}\ee

\noindent therefore, we have that

\be u_{1}=\frac{1-e}{R} \qquad and\qquad
u_{2}=u_{3}=\frac{1+e}{R},\label{ofkasy2}\ee

\noindent are the inverse of the aphelion and the perihelion
distance, respectively. Is possible to solve the roots equation to
obtain the exact value of $u_{4}u_{5}=-\sigma$ (See appendix B for
more details), resulting

\be \sigma=\frac{2}{R}\frac{1-e^{2}}{3-e}.\label{ofkasy3}\ee

\noindent  With this substitution, the equation (\ref{2.1.12}) can
be integrated in terms of elementary functions. Assuming that the
particles are falling from the aphelion distance, where we choose
$\phi=0$, we obtain

\begin{widetext}
\be
\phi=\frac{\sqrt{2(5+3e^{2})(3-e)}}{5-3e}\left[-\ln(\tan\frac{\chi}{4})+
\sqrt{\frac{8e(1-e)}{(5+e)(3-e)}}\arctan\sqrt{\frac{e(3-e)(1+\cos\chi)}{(5+e)(1-e)}}\right].
\label{2.1.11.5}\ee
\end{widetext}

\noindent The second function on the right hand side correspond to
the cosmological correction of the Schwarzschild solution founded
by \cite{Chandrasekhar}.
\medskip

(iii)\underline{\textbf{Circular orbits}}
\medskip

In this case, we have that $r=r_{c}=Cte$, where $r_{c}$ correspond
to the distance of the circular orbit from the singularity. Is
possible calculate the periods of these orbit making use of the
constant of  motion $E$ y $L$, given by (\ref{1.8}) and
(\ref{1.11}) respectively. In fact,  dividing between them, we
find

\be dt=\frac{E}{L}\frac{r_{c}}{f(r_{c})}d\phi, \label{2.1.11.6}\ee

The circular orbit condition, $(V_{eff}^{2})'=0$, give us a five
degrees equation in the variable $r$, moreover, this equation can
be solved for $L$ in term of $r_{c}$, becomes

\be
L=\sqrt{\frac{r_{c}^{5}+\ell^{2}Mr_{c}^{2}}{\ell^{2}(r_{c}-3M)}},\label{2.1.11.7}\ee

\noindent and also, for this orbits $E^{2}=V_{eff}^{2}(r=r_{c})$.
In a period, $\Delta t \equiv T_{t}$, we have $\Delta \phi =2\pi$.
Thus, replacing this in (\ref{2.1.11.6}), we obtain

\be
T_{t}=\frac{2\pi\sqrt{\frac{r_{c}^{3}}{M}}}{\sqrt{1+\frac{r_{c}^{3}}{\ell^{2}M}}}
\equiv
\frac{T_{t,Sch}}{\sqrt{1+\frac{r_{c}^{3}}{\ell^{2}M}}},\label{2.1.11.8}\ee

\noindent where $T_{t,Sch}$ is the period in the coordinate time
founded for the  Schwarzschild spacetime \cite{Shutz}. In the same
way, for the proper time, using equation (\ref{1.11}), yields

\be T_{\tau}=\frac{2\pi
r_{c}\sqrt{\frac{r_{c}}{M}-3}}{\sqrt{1+\frac{r_{c}^{3}}{\ell^{2}M}}}
\equiv \frac{T_{\tau
,Sch}}{\sqrt{1+\frac{r_{c}^{3}}{\ell^{2}M}}},\label{2.1.11.9}\ee

\noindent where $T_{\tau ,Sch}$ is the period in the proper time
founded for the  Schwarzschild spacetime  \cite{Shutz}.

These period are lower than the corresponding to the Schwarzschild
orbits, which is consistent with fact a negative cosmological
constant increase the gravitational attraction.

\subsubsection{\textbf{Orbits Of The Second Kind}}
These type of orbits are allowed when $E$ and $L$ satisfy $E < 1$
and $L^{2} > 12M^{2}$, which are the same conditions of the orbits
of the first kind. Nevertheless, these orbits represents allowed
motions in the regions at the left hand side of the top of the
effective potential. Particles following these orbits fall to the
event horizon.

By simplicity we shall denote with the same name the second kind
trajectories as long as they have the same energy as the orbits of
the first kind.
\medskip

 (i)\underline{\textbf{Planetary orbits}}
\medskip

The planetary orbits of the second kind, posses the same energy
that the planetary orbits of the first kind, moreover these are
starting from an aphelion distance $r=r_{3}=A$, which is greater
than the variable $r$. This mean that the physical range is now
$r<A$ or $u>u_{3}$. Therefore, the cubic polynomial $g(u)$ must be
write as $g(u)=(u-u_{1})(u-u_{2})(u-u_{3})$, and the substitution
is

\be
u=u_{3}+(u_{3}-u_{2})\tan^{2}(\frac{1}{2}\xi),\label{oskpo1}\ee

\noindent where $u_{1}, u_{2}$ and $u_{3}$ are given by equation
(\ref{2.1.11.1}). For this substitution

\beq u=u_{3}\qquad for\qquad \xi=0,\nonumber\\
u\rightarrow\infty \qquad and\qquad r\rightarrow 0 \qquad
for\qquad \xi=\pi.\label{oskpo2}\eeq

Moreover, the form of the solution is the same that in the first
kind counterpart (\ref{2.1.11.3}). Thus

\be \phi = \frac{2}{A}\left(1-\frac{\nu}{\mu+\nu}\right)
F(\xi/2,{\it k})+ \frac{2\nu A}{(\mu+\nu)b}\Pi(\xi/2,{\it k},n),
\label{oskpo3} \ee

\noindent where

\be b=1-4\mu+6\nu+2\mu e.\label{oskpo4}\ee

\be n^{2}=\frac{2(\mu+\nu)}{b}.\label{oskpo5}\ee

\noindent and $A$, ${\it k}$ are given by (\ref{ofkpl1}) and
(\ref{ofkpl2}), respectively.

\bigskip

(ii)\underline{\textbf{Asymptotic orbits}}
\medskip

This case correspond to the continuation of the asymptotic orbit
described above, but now the physical range is $u>u_{2}=u_{3}$.
Therefore, we make the substitution

\be u=\frac{1+e+2e\tan^{2}\frac{1}{2}\xi}{R},\label{oskasy1}\ee

\noindent whit the same value of the roots (\ref{ofkasy2}) and
(\ref{ofkasy3}). Then, after a little manipulation, we obtain

\begin{widetext}
\be
\phi=\frac{\sqrt{2(5+3e^{2})(3-e)}}{5-3e}\left[-\ln(\tan\frac{\chi}{4})+
\sqrt{\frac{8e(1-e)}{(5+e)(3-e)}}\arctan\sqrt{\frac{e(3-e)\sec^{2}\frac{1}{2}\xi}{(5+e)(1-e)}}\right].
\label{oskasy2}\ee
\end{widetext}

In same way that in the asymptotic orbit of the first kind, the
second function on the right hand side correspond to the
cosmological correction to the Schwarzschild solution.

\section{NULL GEODESICS}

\noindent For photons $h=0$, and the effective potential given in
(\ref{1.10}) becomes

\begin{equation}
V^{2}_{eff}=\left(1-\frac{2M}{r}+\frac{r^{2}}{\ell^{2}}\right)\frac{L^{2}}{%
r^{2}}.  \label{3.1}
\end{equation}
\medskip

\begin{figure}[!h]
  \begin{center}
    \includegraphics[width=80mm]{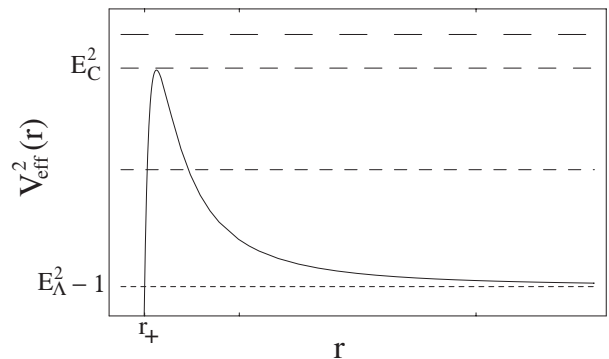}
  \end{center}
  \caption{The figure shows the typical effective potential for non-radial photons. The different level of energy
  are showed.}
  \label{fig:fig.3}
\end{figure}

In FIG.\ref{fig:fig.3} this effective potential has been plotted
for non-radial photons. For different values of the constant $E$
the allowed orbits are as follows
\smallskip

1.- If $E^{2} \leq E^{2}_{\Lambda}-1$, photons plunges into the
singularity from initial distances $r\leq 2M$. For $E^{2} =
E^{2}_{\Lambda}-1$ photons orbits are cardioid (see $IV.2$). This
orbits do not exist in the Schwarzschild case. In the subsection B
we will refer to these orbits as bound orbits.
\smallskip

The following geodesics are unbounded. They are explicitly
integrated in subsection C.
\smallskip

2.- If $E^{2}_{\Lambda}-1 \leq E^{2} \leq E^{2}_{C}$, photons can
fall from the infinity to a minimum distance $r=P$ and them comes
back to the infinity. The photons is only deflected. The other
allowed orbits correspond to photons moving in the other side of
the potential barrier, which plunges into the singularity.
\smallskip

3.- If $E^{2}=E^{2}_{C}$, the photons can orbit in an unstable
circular orbit at $r=3M$. The radius of this orbit is independent
of $L$; $L$ only affects the value of the energy $E^{2}_{C}$.
Photons from the infinity can fall moving asymptotically to the
circle $r=3M$.
\smallskip

4.- If $E^{2}>E^{2}_{C}$, photons from the infinity plunges into
the singularity.
\smallskip

\noindent Therefore, the radial equation in this section is given
by

\begin{equation}
\left(\frac{dr}{d\tau}\right)^{2}= E^{2}-\left(1-\frac{2M}{r%
}+\frac{r^{2}}{\ell^{2}}\right)\frac{L^{2}}{r^{2}}.  \label{3.2}
\end{equation}

\noindent and the effective potential is sketched in
FIG.\ref{fig:fig.3}.

Due to the existence of a negative cosmological constant, the
event horizon, $r_{+}$, is less than $2M$, which value correspond
to the event horizon in the Schwarzschild case. Photons falling
from a distance $r_{0}=2M$, posses a critical energy equal to
$E^{2}_{\Lambda}-1$. Differentiating (\ref{3.2}) with respect to
the affin parameter, $\tau$, we found a {\it pseudo} Newton's law

\be \ddot{r}=-\frac{1}{2}(V_{eff}^{2})'.\label{3.3}\ee

\noindent The {\it radial acceleration} posses a maximum at
$r=4M$, this result is independent of the cosmological constant.
Thus, photons with energy $E^{2}=E^{2}_{C}$ will posses a maximal
radial acceleration, and photons in a intestable circularly orbits
(r=3M, and energy $E_{C}^{2}$) will posses a zero radial
acceleration.
\medskip

On the other side, making the {\it classical} change of variable
$u=1/r$, together with (\ref{1.11}), the radial equation
(\ref{3.2}) can be reduced to

\begin{equation}
\left(\frac{du}{d\phi}\right)^{2}=2Mu^{3}-u^{2}+\frac{1}{D^{2}}=z(u)
\qquad (say), \label{3.4}\end{equation}

\noindent where

\begin{equation}
\frac{1}{D^{2}}=\frac{E^{2}}{L^{2}}- \frac{1}{\ell^{2}}.
\label{3.5}\end{equation}

\noindent denotes the impact parameter. In this case, the impact
parameter is more less than its Schwarzschild counterpart, given
by $\frac{E^{2}}{L^{2}}$.

\subsection{Radial geodesics}

\noindent In this case, $L = 0$, which implies that the radial
equation adopt the simple form

\begin{equation}
\frac{dr}{d\tau}=\pm E,  \label{3.22}
\end{equation}

\noindent where $(+)$ is for outgoing photons and $(-)$ is for
ingoing photons. As in the Schwarzschild case, evaluated in
\cite{Chandrasekhar}, the integration of the above equation yields

\begin{equation}
r=\pm E\tau + r_{0},  \label{3.23}
\end{equation}

\noindent where $r_{0}$ is a constant of integration corresponding
to the initial position of the photon. This result shows that for
photons, in terms of the proper time, the radial motion is
independent of the cosmological constant.

\noindent In order to obtain the equation of motion in terms of
the coordinate time $t$, we makes use of the equations (\ref{1.8}%
) and (\ref{3.22}), yielding

\begin{equation}
\frac{dr}{dt}=\pm(1-\frac{2M}{r}+\frac{r^{2}}{\ell^{2}}).
\label{3.24}
\end{equation}

\noindent or,

\be
\frac{dr}{dt}=\pm\frac{1}{\ell^{2}r}(r-r_{+})(r^{2}+r_{+}r+\rho^{2}).
\label{3.24.1} \ee

\noindent The solution is given by

\begin{widetext}
\be
t=\pm\frac{\ell^{2}r_{+}}{2r_{+}+\rho^{2}}[\ln(r-r_{+})- \frac{1}{2}%
\ln(r^{2}+rr_{+}+\rho^{2})+\delta\arctan(\frac{2r+r_{+}}{\beta})]
+ B, \label{3.25}\ee
\end{widetext}

\noindent where $\alpha=\frac{\rho^{2}}{r_{+}}$, $\beta=\sqrt{%
4\rho^{2}-r_{+}^{2}}$, $\delta=\frac{2\alpha+r_{+}}{\beta}$ , $\rho^{2}=%
\frac{2\ell^{2}M}{ r_{+}}$ and $B$ is a integrating constant.

\subsection{The Bound Geodesics}

\subsubsection{\textbf{The orbits with imaginary eccentricities}}

These type of orbits  can exist when the energy is such that
$0<E^{2}<L^{2}/\ell^{2}\equiv E_{\Lambda}^{2}-1$. Defining a new
parameter $B$ by the expression

\be
\frac{1}{B^{2}}=\frac{1}{\ell^{2}}-\frac{E^{2}}{L^{2}}.\label{ei1}\ee

the radial equation becomes

\be
\left(\frac{du}{d\phi}\right)^{2}=2Mu^{3}-u^{2}-\frac{1}{B^{2}}=2M
h(u), \label{ei2}\ee

\noindent and the physical range is $r_{+}<r\leq r_{3}={\it A}$,
where ${\it A}$ is the aphelion distance for this orbits. The
equation $h(u)=0$ allow only one real root, ${\it A}^{-1}$, and a
complex conjugate pair, $u_{1}$ and $u_{2}$. For this reason, we
shall characterize the roots of the $h(u)$ with a imaginary
eccentricities

\be u_{1}=\frac{-1+ie}{R},\qquad u_{2}=\frac{-1-ie}{R},\qquad
u_{3}=\frac{1}{{\it A}}.\label{ei3}\ee

The sum of the roots of the $h(u)$ allow us to calculate $u_{3}$

\be u_{3}=\frac{1}{{\it
A}}=\frac{1}{2M}+\frac{2}{R}.\label{ei4}\ee

\noindent Thus, the second relation of the roots,
$u_{1}u_{2}+u_{1}u_{3}+u_{2}u_{3}=0$, yields

\beq e^{2}-3=\frac{R}{M}\equiv \frac{1}{\mu}>0,\nonumber\\
\Rightarrow e^{2}>3.\label{ei5}\eeq

Finally, the product of roots, together with (\ref{ei5}), becomes

\be
\left(\frac{R}{B}\right)^{2}=\frac{(1+4\mu)^{2}}{\mu}.\label{ei6}\ee

In this way, the radial equation (\ref{ei2}) is given by

\be \left(\frac{du}{d\phi}\right)^{2}=2M
\left[\left(u+\frac{1}{R}\right)^{2}+\frac{e^{2}}{R^{2}}\right]\left(u-\frac{1}{2M}-\frac{2}{R}\right).
\label{ei7}\ee

Using now the following substitution

\be u=\frac{e \tan\frac{1}{2}\xi -1}{R}.\label{ei8}\ee

Since the range of $u$ is

\be \frac{1}{2M}+\frac{2}{R}\leq u < \infty,\label{ei9}\ee

\noindent the corresponding range of $\xi$ is

\be \xi_{0} \leq \xi < \pi,\label{ei10}\ee

\noindent where

\be \tan\frac{1}{2}\xi_{0}=\frac{6\mu +1}{2\mu e},\label{ei11}\ee

\noindent or, equivalently,

\be \sin\frac{1}{2}\xi_{0}=\frac{6\mu +1}{\triangle} \qquad and
\qquad \cos\frac{1}{2}\xi_{0}=\frac{2\mu
e}{\triangle},\label{ei12}\ee

\noindent where

\be \triangle^{2}=(6\mu +1)^{2}+4\mu^{2}e^{2}.\label{ei13}\ee

We find that with the substitution (\ref{ei8}), the equation
(\ref{ei7}) reduce to

\be \frac{d\xi}{d\phi}=\pm
\sqrt{2\triangle}\sqrt{\sin(\xi-\frac{1}{2}\xi_{0})-\sin\frac{1}{2}\xi_{0}}.\label{ei14}\ee

\noindent Making use of a standard formula in the theory of
elliptic integrals, the solution for $\phi$ can be expressed in
terms of the Jacobian elliptic integral. Thus

\be \pm
\phi=\frac{1}{\sqrt{\triangle}}\int^{\psi}\frac{d\varsigma}{\sqrt{1-k^{2}\sin^{2}\varsigma}},\label{ei15}\ee

\noindent where

\be k^{2}=\frac{1-\sin\frac{1}{2}\xi_{0}}{2},\label{ei16}\ee

\noindent and

\be
\sin^{2}\psi=\frac{1-\sin(\xi-\frac{1}{2}\xi_{0})}{1-\sin\frac{1}{2}\xi_{0}}.\label{ei17}\ee

\noindent From equation (\ref{ei12}) and (\ref{ei17}), it follows
that $\sin^{2}\psi=1$ both when $\xi=\xi_{0}$ (at aphelion) and
$\xi=\pi$ (at the singularity). Moreover, $\sin^{2}\psi=0$ when
$\xi=\tan^{-1}\frac{2\mu e}{6\mu+1}$. Therefore, $\psi$ takes the
value zero within the range (\ref{ei10}) of $\xi$. We conclude
that the range of $\psi$ associated with the range of $\xi$ is

\be -\frac{\pi}{2}\leq \psi \leq \frac{\pi}{2} \qquad (\xi_{0}\leq
\xi \leq \pi).\label{ei18}\ee

\noindent Accordingly, we may write the solution for $\phi$ as

\be
\phi=\frac{1}{\sqrt{\triangle}}[K(k)-F(\psi,k)],\label{ei19}\ee

\noindent where we have assumed that $\phi=0$ at the singularity
where $\xi=\pi$ and $\psi=\pi/2$.

\subsubsection{\textbf{The Cardioid type Geodesic}}

For this case, we shall must consider that $r_{+}\leq r \leq
r_{0}$, and also we suppose that the aphelion distance is
$r_{0}=2M$, this mean that
$E^{2}=\frac{L^{2}}{\ell^{2}}=E_{\Lambda}^{2}-1$, and therefore

\be \dot{r}^{2}=\frac{L^{2}}{r^{3}}(2M-r),\qquad r<2M.
\label{ocf1}\ee

\noindent Dividing the above equation with (\ref{1.11}), we obtain

\be \left(\frac{dr}{d\phi} \right)^{2}=r(2M-r).\label{ocf2}\ee

\begin{figure}[!h]
  \begin{center}
    \includegraphics[width=65mm]{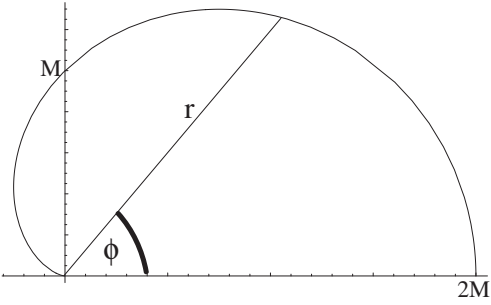}
 \end{center}
  \caption{The figure shows the cardioid  for non-radial photons. The trajectory is independent on the angular
  momentum of the particles and only depend on the mass of black hole.}
  \label{fig:card}
\end{figure}

An elemental integration of the last equation yield

\be r(\phi)=M(1+\cos\phi), \label{ocf3}\ee

which is the equation of the cardioid. A surprising result is the
independence of the angular momentum for the trajectory of non
radial photons in this region. There is only a dependence on the
Black Hole mass.  This trajectory is showed in the
(FIG.\ref{fig:card})

In accordance with our suggest that the photon start from
$r_{0}=2M$, we can say that the photon crosses the event horizon
for an angle $0 < \phi_{+} < \pi$.

\noindent On the other side, replacing (\ref{ocf3}) in
(\ref{1.11}) one get

\be M^{2}(1+\cos\phi)^{2}\frac{d\phi}{d\tau}=L,\label{ocf4}\ee

\noindent After a elemental integration, this last equation
becomes

\be
\tau(\phi)=\frac{M^{2}}{2L}[3\phi+(4+\cos\phi)\sin\phi].\label{ocf5}\ee

\noindent inverting (\ref{ocf3}) and inserting the result in
(\ref{ocf5}) we obtain

\be
\tau(r)=\frac{M^{2}}{2L}\left[3\cos^{-1}\left(\frac{r}{M}-1\right)+\left(3+\frac{r}{M}\right)
\sqrt{2-\frac{r}{M}}\right].\label{ocf6}\ee

\subsection{The unbound geodesics}

\subsubsection{\textbf{The Critical Orbits}}

Returning to the equation (\ref{3.4}), we first consider the
different cases that must be distinguished. They are related with
the disposition of the roots of the cubic equation

\be z(u)=2Mu^{3}-u^{2}+\frac{1}{D^{2}}=0. \label{co1}\ee

\noindent The sum and the products of the roots $u_{1}$, $u_{2}$,
and $u_{3}$ of this equation are given (see appendix B for more
details)

\be u_{1}+u_{2}+u_{3}=\frac{1}{2M} \qquad and \qquad
u_{1}u_{2}u_{3}=-\frac{1}{2MD^{2}}. \label{co2}\ee

\noindent Clearly the equation $z(u)=0$ must allow a negative real
root; and the two remaining roots may either be real (distinct or
coincident) or be a complex conjugate pair. The case when two of
the positive roots coincide plays a specially decisive role in the
discrimination of the null geodesics. Thus, we shall consider this
critical case first.

\noindent The derivative of equation (\ref{co1}), namely,

\be z'(u)=6Mu^{2}-2u=0,\label{co3}\ee

\noindent allows $u=(3M)^{-1}$ as a root; and $u=(3M)^{-1}$ will
be a root of equation (\ref{co1}) (indeed, a double root) if

\be D^{2}=27M^{2} \qquad or \qquad D=3\sqrt{3}M.\label{co4}\ee

\noindent From the condition on the product of the roots, we infer
that the roots of $z(u)=0$ are

\be u_{1}=-\frac{1}{6M} \qquad and \qquad
u_{2}=u_{3}=\frac{1}{3M}.\label{co5}\ee

\noindent Therefore, (\ref{3.3}) can be rewritten as

\be
\left(\frac{du}{d\phi}\right)^{2}=2M\left(u+\frac{1}{6M}\right)\left(u-\frac{1}{3M}\right)^{2}.\label{co6}
\ee

\noindent This equation is satisfied by the substitution

\be
\frac{1}{u}=r=\frac{3M}{\frac{3}{2}\tanh^{2}\frac{1}{2}(\phi-\phi_{0})-\frac{1}{2}},
\label{co7}\ee

\noindent where $\phi_{0}$ is a constant of integration. If
$\phi_{0}$ is chosen to be

\be \tanh^{2}\frac{\phi_{0}}{2}=\frac{1}{3}, \label{co8}\ee

\noindent then

\be u=0 \qquad and \qquad r\rightarrow \infty \qquad when \qquad
\phi=0. \label{co9}\ee

\noindent But we also notice that

\be u=\frac{1}{3M} \qquad when \qquad \phi\rightarrow
\infty.\label{co10}\ee

\begin{figure}[!h]
  \begin{center}
    \includegraphics[width=80mm]{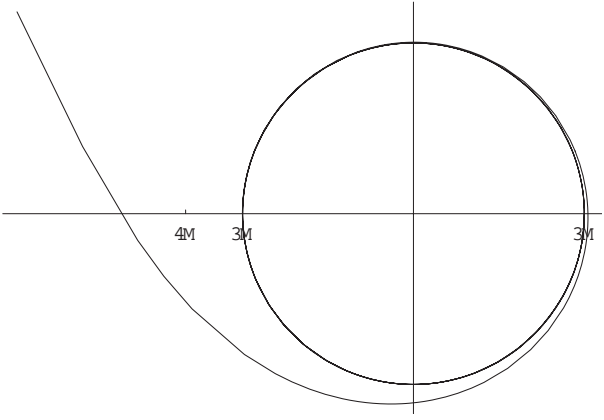}
  \end{center}
  \caption{The polar plot shows the critical  null geodesic.
  The photons arriving from infinite and approaches the circle of radius $3M$, asymptotically, by spiraling
  around it.}
  \label{fig:crit}
\end{figure}

Therefore, a null geodesic arriving from infinite with an impact
parameter $D=3\sqrt{3}M$ approaches the circle of radius $3M$,
asymptotically, by spiraling around it.

Associated with the orbit (\ref{co1}), we must have an {\it orbit
of the second kind} which, originating at the singularity,
approaches, from the opposite side, the same circle at $r=3M$,
asymptotically, by spiralling around it. Such an orbit can be
obtained by the substitution

\be u=\frac{1}{3M}+\frac{1}{2M}\tan^{2}\frac{1}{2}\xi
\label{co11}\ee

\noindent in (\ref{co1}). Then, after a little manipulation it
reduce to

\be
\left(\frac{d\xi}{d\phi}\right)^{2}=\sin^{2}\frac{1}{2}\xi.\label{co12}\ee

\noindent Thus, this last equation is integrating, yields

\be \phi=2\ln(\tan\frac{1}{4}\xi)\qquad or \qquad
\tan\frac{1}{4}\xi=e^{\phi/2}.\label{co13}\ee

\noindent inserting this result in (\ref{co11}), we obtain

\be \frac{1}{r}=u=\frac{1}{3M}+\frac{2
e^{\phi}}{M(e^{\phi}-1)^{2}} \label{co14}\ee

\begin{figure}[!h]
  \begin{center}
    \includegraphics[width=60mm]{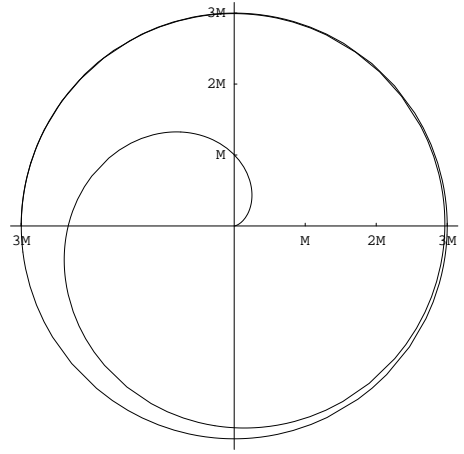}
  \end{center}
  \caption{The polar plot shows the critical  null geodesic of the {\it second kind}. This orbit may be
  considered as a continuation of the critical orbit showed in FIG.\ref{fig:crit}.}
  \label{fig:crit2}
\end{figure}

\noindent along this orbit

\beq u\rightarrow \infty \qquad and \qquad r=0 \qquad when \qquad
\phi\rightarrow 0 \nonumber\\
 and \qquad u=\frac{1}{3M}\qquad as \qquad
\phi\rightarrow \infty.\label{co15}\eeq

\noindent The above solution with the sign of $\phi$ reversed may
be considered as a "continuation" of the solution (\ref{co7}).
\bigskip

(i)\underline{{\it The cone of avoidance}}
\medskip

At any point we can define a {\it cone of avoidance} whose
generators are the null rays, described by the solution
(\ref{co7}) of equation (\ref{co6}), passing through that point,
since, as is clear on general grounds and as we shall establish
analytically in the Orbits With Imaginary Eccentricities below,
light rays, included in the cone, must necessarily cross the
horizon and get trapped.

If $\Psi$ denotes the half-angle of the cone (directed inward at
large distances), then

\be \cot\Psi=+\frac{1}{r}\frac{d\tilde{r}}{d\phi},\label{ca1}\ee

\noindent where

\be
d\tilde{r}=\frac{1}{\sqrt{1-\frac{2M}{r}+\frac{r^{2}}{\ell^{2}}}}dr,
\label{ca2}\ee

\begin{figure}[!h]
  \begin{center}
    \includegraphics[width=65mm]{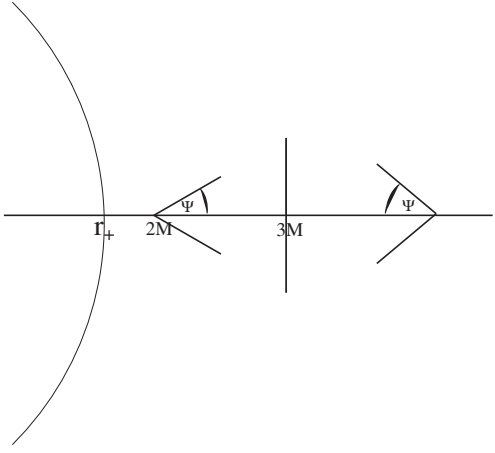}
  \end{center}
  \caption{The figure shows the cones of avoidance. Due to the inclusion of the negative cosmological constant, the
  event horizon appear in a distance lower than $2M$ and
consequently de cone of avoidance at these distance is not close
 as the Schwarzschild case.}
  \label{fig:fig.13}
\end{figure}

\noindent is an element of proper length along the generators of
the cone. Therefore,

\beq
\cot\Psi &=&+\frac{1}{r\sqrt{1-\frac{2M}{r}+\frac{r^{2}}{\ell^{2}}}}\frac{dr}{d\phi}\nonumber\\
&=&-\frac{1}{\sqrt{2M}\sqrt{\frac{1}{2M\ell^{2}}+\frac{u^{2}}{2M}-u^{3}}}\frac{du}{d\phi},
\label{ca3}\eeq

\noindent where $u=1/r$. In equation (\ref{ca3}) we may substitute
for $du/d\phi$ from equation (\ref{co6}). In this way we obtain

\be
\cot\Psi=\frac{\left(\frac{r}{3M}-1\right)\sqrt{\frac{r}{6M}+1}}{\sqrt{\frac{r}{2M}-1+\frac{r^{3}}{2M\ell^{2}}}}
\label{ca4}\ee

\noindent or equivalently

\be
\tan\Psi=\frac{\sqrt{\frac{r}{2M}-1+\frac{r^{3}}{2M\ell^{2}}}}{\left(\frac{r}{3M}-1\right)\sqrt{\frac{r}{6M}+1}}.
\label{ca5}\ee

\noindent From this last equation it follows that

\beq \Psi \sim
\frac{3\sqrt{3}M}{r}\sqrt{1+\frac{r^{2}}{\ell^{2}}}\qquad as\qquad
r\gg 1, \nonumber\\
\Psi=\frac{\pi}{2} \qquad for\qquad r=3M,\nonumber\\
and\qquad \Psi=0\qquad for\qquad r=r_{+}.\label{ca6}\eeq

\noindent where we have assumed that for $r\gg 1$,
$r^{2}/\ell^{2}<1$.

\subsubsection{\textbf{Orbits Of The First Kind}}

This case  occur when all the roots of the cubic equation $z(u)=0$
are real and the two positive roots are distinct. Let the roots be

\be u_{1}=\frac{P-2M-Q}{4MP},\qquad u_{2}=\frac{1}{P},\qquad
u_{3}=\frac{P-2M+Q}{4MP}.\label{ofkn1}\ee

\noindent Notice that the sum of the roots has been arranged to be
equal to $1/2M$ as it is required (see (\ref{B8})), and $u_{1}<0$.
Also, the ordering of the roots, $u_{1}<u_{2}<u_{3}$, requires
that

\be Q+P-6M>0.\label{ofkn2}\ee

Evaluating

\be z(u)=2M(u-u_{1})(u-u_{2})(u-u_{3}),\label{ofkn3}\ee

\noindent with $u_{1}, u_{2}$ and $u_{3}$ given in equation
(\ref{ofkn1}), and comparing the result with the expression
(\ref{3.4}), we obtain the relations

\be Q^{2}=(P-2M)(P+6M),\label{ofkn4}\ee

\noindent and

\be
\frac{1}{D^{2}}=\frac{1}{8MP^{3}}[Q^{2}-(P-2M)].\label{ofkn5}\ee

\noindent Inserting the first of these relation into the second
one get

\be D^{2}=\frac{P^{3}}{P-2M}.\label{ofkn6}\ee

On the other side, with the aid of (\ref{ofkn4}), the inequality
(\ref{ofkn2}) gives

\be (P-2M)(P+6M)>(P-6M)^{2},\label{ofkn7}\ee

\noindent or, after a little manipulation

\be P>3M.\label{ofkn8}\ee

Using the substitution

\be u-\frac{1}{P}=-\frac{Q-P+6M}{8MP}(1+\cos\chi),\label{ofkn9}\ee

\noindent or, equivalently

\be
u+\frac{Q-P+2M}{4MP}=\frac{Q-P+6M}{8MP}(1-\cos\chi),\label{ofkn10}\ee

\noindent we obtain

\beq u=\frac{1}{P}\qquad when \qquad \chi=\pi, \nonumber\\
\bigskip
u=0\qquad and\qquad r\rightarrow \infty \qquad when \nonumber\\
\bigskip
\sin^{2}\frac{\chi}{2}=\frac{Q-P+2M}{Q-P+6M}=\sin^{2}\frac{\chi_{\infty}}{2},\qquad
say\label{ofkn11}\eeq

\noindent substituting (\ref{ofkn9}) and (\ref{ofkn10}) in
(\ref{3.4}) we obtain

\be
\left(\frac{d\chi}{d\phi}\right)^{2}=\frac{Q}{P}(1-k^{2}\sin^{2}\frac{\chi}{2}),\label{ofkn12}\ee

\noindent where

\be k^{2}=\frac{Q-P+6M}{2Q}.\label{ofkn13}\ee

In terms of Jacobi elliptic integrals, our solution can be write
as

\be
\phi=\frac{2P}{Q}\left[K(k)-F(\frac{\chi}{2},k)\right].\label{ofkn14}\ee

\noindent It is possible, in this case, return to original
variable $r$. In fact, using  (\ref{ofkn10}) in the above
equation, we get

\begin{widetext}
\be r(\phi)=\frac{4MP}%
{4Msn^{2}(K(k)-\frac{Q}{2P}\phi)-(Q-P+2M)cn^{2}(K(k)-\frac{Q}{2P}\phi)}.
\label{ofkn15}\ee
\end{widetext}

\begin{figure}[!h]
  \begin{center}
    \includegraphics[width=80mm]{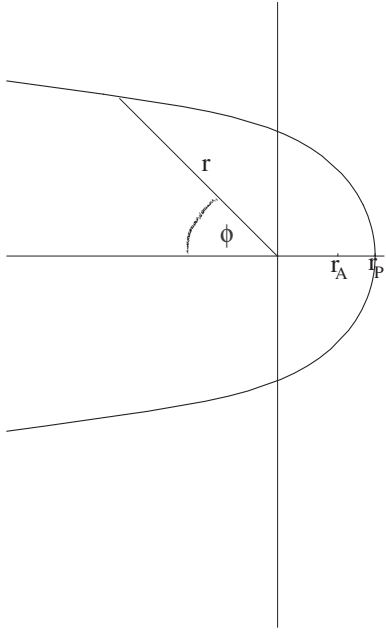}
  \end{center}
  \caption{The polar plot shows the  null geodesic of the {\it first kind}. This orbit correspond to the motion
  of photon, which posses energy $E^{2}_{0}$, such that $E^{2}_{\Lambda}-1 <E^{2}_{0} < E^{2}_{C}$. The motion is unbounded
  and the photon has a perihelion distance, $r=P$.}
  \label{fig:ofk}
\end{figure}

\subsubsection{\textbf{Orbits Of The Second Kind}}

In this case the physical range is $u>u_{3}>u_{2}$. With the
substitution

\be
u=\frac{1}{P}+\frac{Q+P-6M}{4MP}\sec^{2}\frac{\chi}{2},\label{oskn1}\ee

\noindent we obtain

\beq aphelion\qquad u=u_{3}=\frac{Q-P-2M}{4MP} \qquad and \qquad
\chi=0 \nonumber\\
u\rightarrow\infty\qquad and \qquad r\rightarrow 0\qquad
when\qquad \chi=\pi.\nonumber\\
\label{oskn2}\eeq

\noindent Replacing (\ref{oskn1}) in (\ref{3.4}) one get that its
reduce to the same form (\ref{ofkn12}) and with the same value of
$k^{2}$; and we may now write

\be \phi=\frac{2P}{Q}F(\frac{\chi}{2},k).\label{oskn3}\ee

\noindent Therefore, one get

\be r(\phi)=\frac{4MP}%
{4M+(Q+P-6M)nc^{2}(\frac{Q}{2P}\phi)}. \label{oskn4}\ee

\begin{figure}[!h]
  \begin{center}
    \includegraphics[width=60mm]{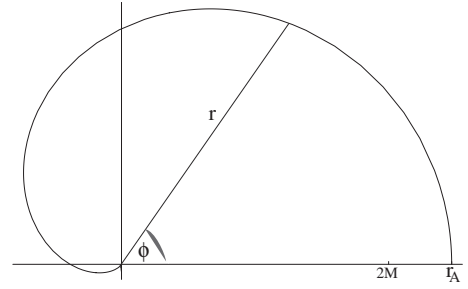}
  \end{center}
  \caption{The polar plot shows the null geodesic of the {\it second kind}. This orbit correspond to the motion
  of photon, which posses energy $E^{2}_{0}$, such that $E^{2}_{\Lambda}-1 <E^{2}_{0} < E^{2}_{C}$. It are falling
  from an aphelion distance, $r=A$, crossing the event horizon, $r_{+}<2M$, and finally arrive to the singularity.}
  \label{fig:crit2}
\end{figure}

\subsubsection{\textbf{Orbits With Imaginary Eccentricities}}

 Finally, we consider the case when two roots of the equation
 $z(u)=0$ are complex conjugate pair, i.e. $u_{2}=u^{\ast}_{3}$, while the remained root has a
 negative value. We shall now write the roots in term of an
 imaginary eccentricities as

 \be u_{2}=\frac{1+ie}{R}\qquad and \qquad
 u_{3}=\frac{1-ie}{R}.\label{ein1}\ee

\noindent in this way, using (\ref{B8}) we get for the negative
real valued root $u_{1}$

\be u_{1}=\frac{1}{2M}-{2}{R},\label{ein2}\ee

\noindent yielding to the following expression for $z(u)$

 \be
 z(u)=2M\left(u-\frac{1}{2M}+\frac{2}{R}\right)\left[\left(u-\frac{1}{R}\right)+\frac{e^{2}}{R^{2}}\right],
 \label{ein3}\ee

The sum of products of the roots, (\ref{B9}), and the product of
the root, (\ref{B10}), gives

\be e^{2}=\frac{3\mu-1}{\mu},\label{ein4}\ee

\noindent and

\be \frac{D^{2}}{M^{2}}=\frac{1}{\mu(4\mu-1)^{2}},\label{ein5}\ee

\noindent respectively. These relation requires

\be \mu > \frac{1}{3}\qquad and \qquad D <
3\sqrt{3}M.\label{ein6}\ee

The substitution is

\be u=\frac{1+e\tan\frac{\xi}{2}}{R},\label{ein7}\ee

\noindent and the range of $u$ is now

\be \infty>u>u_{0}, \qquad \pi>\xi>\xi_{0}.\label{ein8}\ee

Since these orbits are unbounded, the solution for $\phi$ is given
by the same equation (\ref{ei15}), but now $e$ have another value.
In particular

\be
\phi_{\infty}=\frac{1}{\sqrt{\triangle}}[K(k)-F(\psi_{\infty},k)],\label{ein9}\ee

\noindent where $\psi_{\infty}$ (on substituting for $e^{2}$ its
present value $\frac{3\mu-1}{\mu}$) is now given by

\be
\sin^{2}\psi_{\infty}=\frac{\triangle+1}{\triangle+6\mu-1},\label{ein10}\ee

\noindent where it should remembered that $\mu >1/3$; and further
that

\be \triangle=\sqrt{48\mu^{2}-16\mu+1}.\label{ein11}\ee

\section{Conclusion}
We have studied the geodesic structure of AdS black holes
analyzing the behavior of null and time-like geodesic by means of
the effective potential which appears in the radial equation of
motion.

For radial time-like geodesics, the effective potential is
unbounded since $\Lambda < 0$ acts as an attractive force which
increases with the distance to the horizon. Particles always
plunges into the horizon from an upper distance determined by the
constant of motion $E$, which means that all radial particles are
confined.

For non-radial time-like geodesics bounds orbits are allowed if
$L^{2}>11.25M^{2}$. For $L^{2}=11.25M^{2}$ there is an unstable
circular orbits. For $L^{2}<11.25M^{2}$ no bounded orbits are
allowed.

We have explicitly evaluated the proper and coordinate time in
terms of the radial coordinates for radial time-like geodesics.
Due to the attractive nature of the negative cosmological
constant, these times are lower than in the Schwarzschild case.
Moreover, Due to the Schwarzschild case is possible have
non-confined particles, this comparatione is valid when the
confined particles are considered. In this range, we get that
physical features are the same in both cases.

We found that if $E < 1$ and $L^{2}>11.25M^{2}$, exist the orbits
of the first kind which represents bound geodesics, with two
extreme values. Three classes of orbits can be identify in this
case: planetary orbits, circular orbits and asymptotic orbits,
which approaches to the circular orbits by spiraling around it.
Orbits of the second kind posses the same energy that the orbits
of the first kind, but exist the regions at the left hand side of
the top of the effective potential.

In the planetary orbits described above, we found an expression to
the advance of perihelion in term of an infinity serial including
the cosmological constant contribution, which contrast with the
works of G. Kraniotis and S. Whitehouse \cite{K-W}, where the
evaluation was done by means of the genus $2$ siegelsche modular
forms and including the Mercury's data. Our advantage consist in
recuperate the classical serial for the Schwarzschild case,
founded firstly by Einstein, for vanished cosmological constant.

For radial null geodesics, we obtain that, in the proper time, the
situation is the same as in the Schwarzschild case. For the
coordinate time we obtain a term which represent the Schwarzschild
solution, founded in \cite{Chandrasekhar}, plus the cosmological
constant contribution.

The effective potential for non-radial null geodesics show that
the bounded orbits are allows if the constants of motion satisfies
the condition $E^{2} < L^{2}/\ell^{2}$. These type of orbit not
exist in the Schwarzschild spacetime, and correspond to confined
photons which posses a return distance and plunges to the event
horizon. The analytic solution is expressed by mean of  elliptic
function of Jacobi. If $E^{2} = L^{2}/\ell^{2}$, the return
distance is $r=2M$, the event horizon of the Schwarzschild
spacetime, which correspond to a new type the orbit no allowed
before. The analytic solution correspond to the {\it cardioid}
type geodesic. If $E^{2} > L^{2}/\ell^{2}$, the exterior AdS black
hole spacetime have the same form that the Schwarzschild
spacetime. The cosmological constant effects are showed by mean of
the cone of avoidance, which show that the photons feel an {\it
attractive force} and the spacetime result to be more greater than
the Schwarzschild spacetime. The likeness of these spacetimes
implies that the solutions have the same analytic form, as we was
showed in this paper (Section III.c).

\begin{acknowledgments}

We acknowledge fruitful discussions with members of the GACG
(www.gacg.cl), specially with S. Lepe and M. Valenzuela. Also, we
acknowledge useful discussions with M. Plyushchay, A.
Anabal\'{o}n, E. Carqu\'{\i}n and P. Landeros.  (M.O) and (J.R.V)
thank to USACH for hospitality. This work was supported by CONICYT
through Grant FONDECYT No. 1040229 (NC).

\end{acknowledgments}

\appendix
\section{Solution to the cubic equation $ x^{3}+ax-b=0 $.}

We are interest in solve cubic equations of the form

\be x^{3}+ax-b=0.\label{A1}\ee

\noindent where $a>0$ and $b>0$ (see equation (\ref{1.3}), for
instance). For this purpose, we make the change of variable
$x=Z\sinh\theta$, and multiplying by a scalar $\alpha$, we  obtain
the following equation

\be \alpha Z^{3}\sinh^{3}\theta + \alpha a Z\sinh\theta-\alpha
b=0.\label{A2}\ee

\noindent Considering the hyperbolic identity

\be
4\sinh^{3}(\theta)+3\sinh(\theta)-\sinh(3\theta)=0.\label{A3}\ee

\noindent and comparing (\ref{A2}) with (\ref{A3}), we obtain the
following relations

\be \alpha Z^{3}=4,\qquad \alpha a Z=3,\qquad \alpha
b=\sinh(3\theta).\label{A4}\ee

\noindent Solving the above equation for $Z$ and $\theta$, we
obtain

\be Z=2\sqrt{\frac{a}{3}},\label{A5}\ee

\be \theta=\frac{1}{3}\sinh
^{-1}\left(\frac{3b}{2}\sqrt{\frac{3}{a^{3}}}\right)+\frac{2\pi}{3}n
i,\label{A6}\ee

\noindent where the period of the hyperbolic function is $2\pi i$.
Therefore, the roots of the cubic equation are obtain replacing
$n=0,1,2$. Thus, the roots of (\ref{A1}) are

\be x_{1}=2\sqrt{\frac{a}{3}}\sinh\left[\frac{1}{3}\sinh
^{-1}\left(\frac{3b}{2}\sqrt{\frac{3}{a^{3}}}\right)
\right],\label{A7}\ee

\be x_{2}=2\sqrt{\frac{a}{3}}\sinh\left[\frac{1}{3}\sinh
^{-1}\left(\frac{3b}{2}\sqrt{\frac{3}{a^{3}}}\right)+\frac{2\pi}{3}
i \right],\label{A8}\ee

\be x_{3}=2\sqrt{\frac{a}{3}}\sinh\left[\frac{1}{3}\sinh
^{-1}\left(\frac{3b}{2}\sqrt{\frac{3}{a^{3}}}\right)+\frac{4\pi}{3}
i \right].\label{A9}\ee

\noindent Thus, considering (\ref{1.3}), the event horizon $r_{+}$
is obtained like (\ref{1.4}).

\section{Roots of the polynomials $\texttt{P}(u)$}

We are interest in rewrite $5$ degrees  polynomials from it
standard form

\be \texttt{P}(u)=u^{5}-a u^{4}+b u^{3}-c u^{2}+d
u-e,\label{B1}\ee

\noindent to it factorized form

\be
\texttt{P}(u)=(u-u_{1})(u-u_{2})(u-u_{3})(u-u_{4})(u-u_{5}),\label{B2}\ee

\noindent where the $u_{i}$ ($i=1,...,5$) are the roots of
$\texttt{P}(u)$. This roots must satisfies the following relations

\be u_{1}+u_{2}+u_{3}+u_{4}+u_{5}=a,\label{B3}\ee

\beq
u_{1}u_{2}+u_{1}u_{3}+u_{1}u_{4}+u_{1}u_{5}+u_{2}u_{3}+u_{2}u_{4}\nonumber\\
+u_{2}u_{5}+u_{3}u_{4}+u_{3}u_{5}+u_{4}u_{5}=b,\label{B4}\eeq

\beq
u_{1}u_{2}u_{3}+u_{1}u_{2}u_{4}+u_{1}u_{2}u_{5}+u_{1}u_{3}u_{4}\nonumber\\
+u_{1}u_{3}u_{5}+u_{1}u_{4}u_{5}+u_{2}u_{3}u_{4}+u_{2}u_{4}u_{5}\nonumber\\
+u_{2}u_{3}u_{5}+u_{3}u_{4}u_{5}=c,\label{B5}\eeq

\beq
u_{1}u_{2}u_{3}u_{4}+u_{1}u_{2}u_{3}u_{5}+u_{1}u_{2}u_{4}u_{5}\nonumber\\
+u_{1}u_{3}u_{4}u_{5}+u_{2}u_{3}u_{4}u_{5}=d,\label{B6}\eeq

\be u_{1}u_{2}u_{3}u_{4}u_{5}=e.\label{B7}\ee

If we have a n degrees polynomial, with $n<5$, then the above
formulas can be used. For example if $n=3$, implies that $d=e=0$
and $u_{4}=u_{5}=0$, therefore the two last equations,(\ref{B6})
and (\ref{B7}) are equals to zero (i.e.,
$\texttt{P}(u)=u^{2}(u^{3}-au^{2}+bu-c)\equiv
u^{2}\texttt{g}(u)$). The remaining equations are

\be u_{1}+u_{2}+u_{3}=a,\label{B8}\ee

\be u_{1}u_{2}+u_{1}u_{3}+u_{2}u_{3}=b,\label{B9}\ee

\be u_{1}u_{2}u_{3}=c.\label{B10}\ee


\end{document}